# THE LIGO INTERFEROMETER SENSING AND CONTROL SYSTEM

J. Heefner, R. Bork, R. Abbott, LIGO Laboratory, California Institute of Technology, Pasadena, CA 91125, USA

Abstract

The LIGO Interferometer Sensing and Control System (ISC) is a large and highly distributed Multiple Input Multiple Output (MIMO) control system that is used to control the length and alignment degrees of freedom of the interferometers. The 4 kilometer Fabry-Perot cavity lengths are controlled to better than $10^{-13}$ meters (rms) and the angular degrees of freedom are controlled to better than $10^{-8}$ radians. This paper will describe the real-time digital servo control systems that have been designed, developed and implemented for the LIGO Length Sensing and Control (LSC) [1] and Alignment Sensing and Control (ASC) [2] systems. In addition, the paper will describe how these controls, along with the suspended optic controls [3], have been integrated into the overall LIGO control and data acquisition system [4].

## 1 INTRODUCTION

The LIGO interferometers located in Hanford, Washington and Livingston, Louisiana are Michelson laser interferometers enhanced by multiple coupled optical resonators. These coupled optical resonators are 4 kilometer long Fabry-Perot cavities in each arm of the interferometer. The mirrors that form the cavities are suspended from a single loop of wire mounted inside suspension cages that are, in turn, mounted on seismically isolated optical platforms within the LIGO vacuum system. Control of the optic is achieved using voice coil actuators that act on magnets attached to the surface of each optic. Shadow sensors are used to measure movement and orientation of the optic with respect to the suspension cage. This "local" damping of each optic is not sufficient to keep the optical cavities of the interferometer on resonance in the presence of disturbances. It is the function of the LSC and ASC systems to actively control the length and alignment of the optical cavities with respect to the "on resonance" condition. Figure 1 shows the LSC photodetectors and ASC wavefront sensors (WFS) and Quadrant photodetectors (QPDX, QPDY) in relation to the interferometer optical components.

There are many components that are common to both the LSC and ASC. The length and alignment degrees of freedom are detected and controlled by impressing phase modulated radio frequency sidebands on the light injected into the interferometer. Tuned photodetectors whose outputs are band pass filtered (BPF) and synchronously demodulated are used to detect the various degrees of freedom. The demodulated output is then low pass filtered (LPF) passed through a whitening filter, anti-alias filter (AA) and on to an analog to digital converter (ADC). A "typical" input signal path is shown in figure 2.

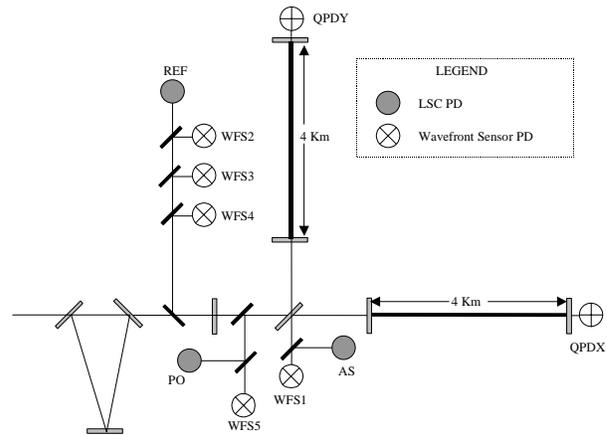

Figure 1: LSC and ASC System Layout

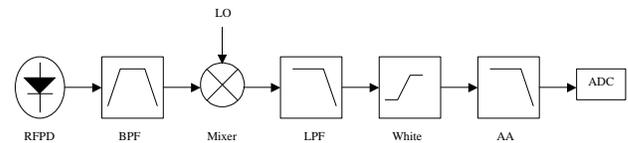

Figure 2: "Typical" Input Signal Path

Outputs from the LSC and ASC systems are handled by the suspension control system as described in [3]. The interface to the suspension controls is via fiber optic reflective memory links. There are three reflective memory loops for the ASC and LSC controls, one for each arm of the interferometer and one for the corner station. The loop to each arm is used to pass data between front end systems in the corner station and each end station. A loop is used for each arm in order to minimize and balance the 13-microsecond delay caused by the speed of light transit time of the 4-kilometer arms. The loop in the corner station is used to pass data between LSC, ASC and suspension front end CPUs located at the interferometer vertex. In addition there is a reflective memory network for the LIGO data acquisition system. Each node of a particular reflective memory network is connected in a loop using single mode or multi-mode fiber. Data written to a memory location at one node is passed to each node in the loop. In this way data can be shared among nodes.

The analog electronics are very low noise (1-2nV/√Hz input referred) and housed in 6U height Eurocard format crates in the ISC control racks. All ADCs, DACs, processors and other commercial equipment are VME. The timing for the ADCs and DACs is derived from the same GPS (Global Positioning System) based timing system used by the LIGO data acquisition system [4].

## 2 LENGTH SENSING AND CONTROL

Figure 3 is a block diagram of the length control signal paths from each of the LSC photodetectors to each of the interferometer optics. As can be seen for the figure the length control system is a highly distributed multi-input, multi-output (MIMO) servo control system.

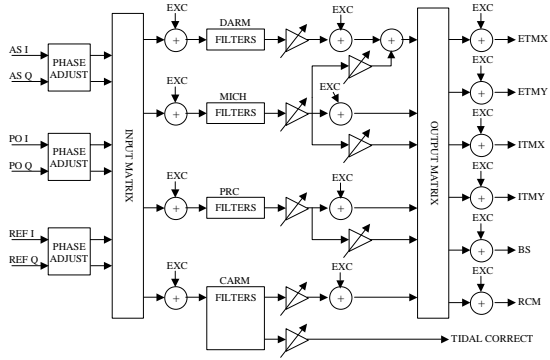

Figure 3: Length Controls showing signal paths from anti-symmetric, pick-off and reflected photodetectors (AS, PO, REF, respectively) to each optic.

The in-phase (I) and quadrature-phase (Q) component of each photodetector are demodulated and passed through a software block that can offset the overall phase of each signal. These phase adjusted signals are passed to the input matrix where each of the degrees of freedom to be controlled are calculated. These degrees of freedom are differential arm length (DARM), michelson length (MICH), power recycling cavity length (PRC) and common arm length (CARM), shown top to bottom, respectively in figure 3. Each degree of freedom is then filtered. The filters shown in figure 3 represent a block of code the can contain up to 10 infinite impulse response (IIR) filters. Each of these filters can contain up to 4 second order sections and each filter can be enabled or bypassed on the fly. Coefficients for each of the filters are read from a modifiable text filter on system startup, but an effort is currently under way to allow the coefficients file to be read and reloaded without restarting the system. Following the filters the signal is passed to an operator adjustable gain stage and then on to the output matrix where the appropriate actuation signals are produced. These actuation signals are passed via reflective memory to the digital suspension controls, which then add the signals to the appropriate degrees of freedom of each optic (ETMX, ETMY, ITMX, ITMY, BS and RCM) as described in [3]. Also shown in figure 3 are the points at which GDS (Global Diagnostic System) excitation signals (EXC) can be added into the signal path. These signals are used by the operators to measure transfer functions, and run other diagnostic tests as described in [4].

Not shown in figure 3 is the connection between the real-time servo calculations and the interferometer lock acquisition code [5]. The lock acquisition code looks at data from the system and determines in real time the filters, gains, etc. required to bring the optical cavities of the interferometer into resonance.

All sampling and servo calculations for the LSC are performed at 16,384Hz, the maximum sample rate for LIGO. This sample rate is necessary because many of the LSC servos have unity gain bandwidths greater than 100Hz. In addition, data collected and processed by a CPU located in the vertex area may be used to drive an optic located in one of the end stations. For these situations, the 13 microseconds required to transit the 4 kilometer arms, must be taken into account when the servos are designed.

## 3 ALIGNMENT SENSING AND CONTROL

Figure 4 is a block diagram of the alignment control signal paths from each of the ASC wavefront sensor photodetectors (WFS) to each of the interferometer optics. As can be seen from the figure, the ASC, like the LSC, is a highly distributed MIMO servo control system. Not shown in the figure are the connections from each of the QPDs located in the end stations to steering optics located in the interferometer vertex.

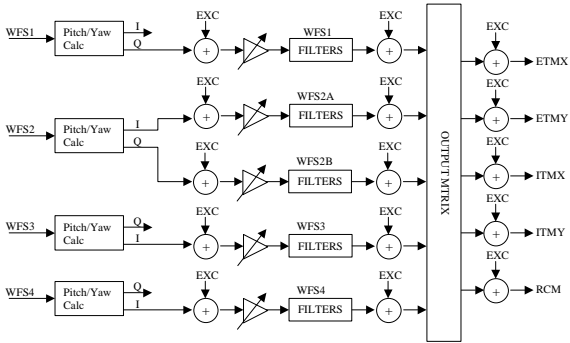

Figure 4: Alignment controls showing signal paths from each wavefront photodetector to each optic

Each of the WFS and QPD photodetectors are a 4 segment detector. Each segment of the detector is processed individually and then used to determine beam pointing (pitch and yaw). One difference between the WFS and the QPD is that the WFS is sensitive to the RF sideband modulation impressed on the interferometer light and the QPD is sensitive to the carrier.

As can be seen in figure 4, the in-phase (I) and quadrature-phase (Q) of each WFS are detected and used to determine an I and Q phase pitch and yaw signal for that WFS. Then, depending on which WFS it is, the I and/or Q pitch and yaw signals are passed to operator adjustable gain stages and then to filters. The filters shown in figure 4 represent blocks of code that can contain up to 10 IIR filters and each of these filters can contain up to 4 second order sections. Each filter can be enabled or bypassed on the fly and like the LSC the coefficients for each of the filters are read from a text file on system startup. Following the filters, signals are passed to an output matrix where the actuation signals for each of the interferometer optics are calculated. These actuation signals are then passed to the digital suspension controls via reflective memory where they are applied to the appropriate degrees of freedom as described in [3].

One difference between the ASC and LSC is that data for the servos is collected in both the end stations and the vertex. Data collected from the QPDs in the end stations is passed to the vertex via the reflective memory network for the respective arm and all of the calculations for the beam pointing servo are done by the CPU in the vertex. Also shown in figure 4 are the points at which GDS excitation signals (EXC) can be added into the signal path. All sampling and servo calculations for the ASC are performed at 2048Hz.

## 4 INTERFACE TO DAQ, GDS AND EPICS

The LIGO DAQ and GDS systems are interfaced via reflective memory. One of the functions of the vertex input crate is to read and write data from and to this reflective memory. Using the DAQ and GDS an operator can look at data from the system, measure transfer functions, plot power spectra and input diagnostic test signals, in real time. A more complete description of the functions and operation of the DAQ and GDS systems are described in reference [4].

Routine operator interface is via EPICS[1]. Through EPICS, the operator can monitor the system, change matrices, gains and offsets, engage filters and invert servo polarity. A Motorola MVME 162 CPU is located in each of the LSC and ASC front end VME crates. This CPU processes the database records, state code, etc. for the controls located in that crate. Data is communicated between the front end CPU (Pentium III) and the EPICS CPU via shared memory across the VME backplane. This is done in an effort to minimize the burden on the front end CPU and allow it to handle only the time critical front end servo controls.

## 5 ACKNOWLEDGEMENTS

We thank the entire LIGO team for assistance and support. This work was supported by National Science Foundation Grant PHY-920038.

---

[1] EPICS- Experimental Physics and Industrial Control System